\documentclass{bmcart}

\usepackage{amsthm,amsmath}
\RequirePackage{natbib}
\RequirePackage{hyperref}
\usepackage[utf8]{inputenc} % unicode support

\usepackage{multirow}

\usepackage{graphicx} % Include figure files
\graphicspath{{.}{./Figs/}}

\startlocaldefs

\endlocaldefs

\begin{document}

%%% Start of article front matter
\begin{frontmatter}

\begin{fmbox}
\dochead{Research}

\title{Community structure in co-inventor networks affects time to first
citation for patents}

\author[
   addressref={aff1}
]{\inits{W}\fnm{William} \snm{Doonan}}
\author[
   addressref={aff1},
   noteref={n1}
]{\inits{KW}\fnm{Kyle W} \snm{Higham}}
\author[
   addressref={aff1}
]{\inits{M}\fnm{Michele} \snm{Governale}}
\author[
   addressref={aff1}, % id's of addresses, e.g. {aff1,aff2}
   corref={aff1},     % id of corresponding address, if any
%   noteref={n1},     % id's of article notes, if any
   email={uli.zuelicke@vuw.ac.nz} % email address
]{\inits{U}\fnm{Ulrich} \snm{Z{\"u}licke}}

\address[id=aff1]{% unique id
  \orgname{Te P{\=u}naha Matatini, School of Chemical and Physical
           Sciences, Victoria University of Wellington}, % university, etc
  \street{PO Box 600}, %
  \postcode{6140} % post or zip code
  \city{Wellington}, % city
  \cny{New Zealand} % country
}

\begin{artnotes}
%\note{Sample of title note} % note to the article
\note[id=n1]{Present address: College of Management of Technology,
EPFL, Odyssea, Station 5, 1015 Lausanne, Switzerland} % note, connected to author
\end{artnotes}

\end{fmbox} % comment this for two column layout

\begin{abstractbox}

\begin{abstract} % abstract
%\parttitle{First part title} % if any
We have investigated community structure in the co-inventor network of a
given cohort of patents and related this structure to the dynamics of how
these patents acquire their first citation. A statistically significant
difference in the time lag until first citation is linked to whether or not
this citation comes from a patent whose listed inventors share membership
in the same communities as the inventors of the cited patent. Although the
inventor-community structures identified by different community-detection
algorithms differ in several aspects, including the community-size
distribution, the magnitude of the difference in time to first citation is
robustly exhibited. Our work is able to quantify the expected acceleration
of knowledge flow within inventor communities and thereby further
establishes the utility of network-analysis tools for studying innovation
dynamics.
\end{abstract}

%%%%%%%%%%%%%%%%%%%%%%%%%%%%%%%%%%%%%%%%%%%%%%%%%%%%%%%%%%%%%%%%%%%%%%%%%%%
%%                                                                       %%
%% The keywords begin here. Put each keyword in separate \kwd{}.         %%
%%                                                                       %%
%%%%%%%%%%%%%%%%%%%%%%%%%%%%%%%%%%%%%%%%%%%%%%%%%%%%%%%%%%%%%%%%%%%%%%%%%%%


% MSC classifications codes, if any
%\begin{keyword}[class=AMS]
%\kwd[Primary ]{}
%\kwd{}
%\kwd[; secondary ]{}
%\end{keyword}

\end{abstractbox}
%
%\end{fmbox} % uncomment this for twcolumn layout

\end{frontmatter}

%%% start of article main body
% <put your article body there>

\section*{Introduction and Motivation}

Inventions can be codified in patent applications that, if granted, bestow
certain rights to the assignees. Patent documents contain citations to
other patents as part of the requirement to acknowledge the state of prior
art and to delimit the invention's legal scope. These citations are
understood to represent knowledge flows between inventors and have been
used widely to study various economic and social aspects of innovation
dynamics~\cite{hal02,bre04,jaf17}.

The degree to which social and working relationships between inventors
influence patenting and knowledge propagation in technological-innovation
space has been the subject of recent interest~\cite{sto02,bal04,sin05,
sor06,bre09,bre16,tot18}. We shed new light on this topic from a
network-analysis perspective by studying the effect of community structure
in the co-inventor network on patent-citation dynamics. It can be
intuitively expected that knowledge about inventions will be transmitted,
and thus become available for utilization, faster within groups of
inventors that have collaborated before. One of the main aims of our
present work is to verify this expectation and quantify the acceleration
of knowledge flow through inventor communities. We use the time lag until
first citation~\cite{gay05,lee17} as a proxy measure for the fastest speed
at which knowledge about an invention can propagate. This is motivated by
the fact that, as in the case of an electrical pulse whose leading edge is
the real carrier of information, independent of the rest of its line shape,
the time to first citation is more representative of the speed of
information flow than any other, aggregate or average, citation measure.

In previous studies, inventor communities were identified using attributes
such as geographical co-location~\cite{jaf93,sin05,bre16} or employment at
the same firm~\cite{kog92,sto02,sin05,bre09}. Here our approach is
different. We use established community-detection algorithms~\cite{for10,
for16} to identify communities of inventors in the co-inventor network that
has been constructed by projection from a bipartite inventor-and-patent
network. This method can directly reveal communities based on collaborative
inventive activity without relying on any externally observed relationship
proxies. On the other hand, the algorithmic identification of communities
may introduce biases arising from idiosyncrasies inherent to particular
detection methods. Our present work also investigates how robustly
community effects in patent-citation dynamics can be identified using the
generally different community structures obtained by various established
detection algorithms.

Community structures in citation networks for scientific articles have been
analyzed recently (see, for example, Refs~\cite{lam09,che10,liu17}), and
opportunities for similar studies on the available large amounts of rich
patent-citation data are now also beginning to be realized~\cite{nak18,
gao18}. Our study offers an example for the type of interesting insights
that can be gained by applying network-analysis-based community-detection
methods in the important area of technical innovation.

The remainder of this article is structured as follows. In the following
\emph{Methods\/} section, we start by discussing the data used to construct
the co-inventor network and the five community-detection algorithms
utilized in our work. Basic properties of communities identified on the
largest connected component of the co-inventor network by the different
algorithms are compared before details are given on how the citation-lag
distributions are assembled. The subsequent \emph{Results and Discussion\/}
section presents a detailed analysis of the observed distributions for the
time to first citation, focusing in particular on how these depend on
whether inventors from the citing and the cited patent share membership in
one of the algorithmically detected communities. Our \emph{Conclusions\/}
are summarized in the final section. A list of abbreviations used
throughout this article is given in Table~\ref{tab:abbrevs}.

\section*{Methods}

The availability of disambiguated US-patent data~\cite{li14} makes it
possible to construct the co-inventor network associated with a particular
cohort of patents. To be specific, we use patents granted by the US Patent
and Trademark Office (USPTO) during 1995-1999 and assigned to one of
the following classes from the US Patent Classification System
(USPC)~\cite{USPC}: 257 (Active solid-state devices), 326 (Electronic
digital logic circuitry), or 438 (Semiconductor device manufacturing:
process)\footnote{The objectives of a particular investigation will
generally determine the choice of cohort. Ours is motivated by three main
needs: (i) to leave a sufficiently long time window for citation accrual,
(ii) to have a sufficiently large data set to reduce statistical noise, and
(iii) to include patents from similar fields that are granted around the
same time to minimize variations in the inventors' work environment.}. The
projection of the bi-partite patent-and-inventor network onto inventors
yields the co-inventor network~\cite{bal04} associated with this cohort.
Going further than using a simple projection as was done in previous
work~\cite{bal04,sin05}, we introduce edge weights $w_{ij}$ reflecting the
frequency and intensity of the co-inventing activity~\cite{new01} between
two inventors $i$ and $j$;
\begin{equation}\label{eq:weights}
w_{ij} = \sum_\alpha \frac{\delta_i^{(\alpha)}\,
\delta_j^{(\alpha)}}{n_\alpha - 1} \quad .
\end{equation}
The sum in Eq.~(\ref{eq:weights}) is over all patents in the chosen cohort
that list more than one inventor, and $\delta_i^{(\alpha)} = 1$
($\delta_i^{(\alpha)} = 0$) if inventor $i$ is (not) among the $n_\alpha$
inventors listed on the particular patent $\alpha$. In the following, we
restrict ourselves to considering only the largest connected component
(LCC) of the thus-constructed co-inventor network, which can be expected to
capture the most relevant aspects of inventors' connectedness~\cite{new01a,
bre09}. Amounting to approximately 31\% of the total co-inventor network by
number of nodes, the LCC comprises a diverse assortment of inventors as
evidenced by the spread of firms their patents are assigned to. For
example, IBM, Hitachi LTD, Motorola INC, Kabushiki Kaisha Toshiba, and
Texas Instruments represent $14$\%, $7.2$\%, $4.6$\%, $4.1$\%, and $4.0$\%
of nodes in the LCC, respectively. Other companies each account for
portions smaller than $4$\% of the total in the component.
Table~\ref{tab:summStat} provides an overview of relevant summary
statistics pertaining to the patent cohort and co-inventor network
considered here.

We have used five established community-detection algorithms to analyze the
LCC of the co-inventor network: Greedy~\cite{new04a}, Louvain~\cite{blo08},
Infomap~\cite{ros08}, Random Walks~\cite{pon06}, and Propagating
Labels~\cite{rag07}. Although different algorithms generally yield
different community structures, clear similarities are exhibited between
the structures obtained by conceptually related approaches such as Greedy
and Louvain on the one hand, and Infomap, Random Walks, and Propagating
Labels on the other. In particular, we observe the well-known
resolution-limit issue~\cite{for07} where the communities delineated by
approaches that maximise modularity~\cite{new04} (Greedy and Louvain) are
typically larger in size and generally subsume the many smaller communities
identified by other approaches (Infomap, Random Walks, Propagating Labels).
This may be inferred graphically from visualizations of community
structures, such as those shown in Fig.~\ref{fig:LCCplots}. A more
quantitative comparison is possible based on the size distributions for
communities obtained by application of each of the five different detection
algorithms to the LCC, given in Fig.~\ref{fig:commDist}, and the
corresponding community-structure-related summary statistics provided in
Table~\ref{tab:commStat}. We also use the adjusted Rand index
(ARI)~\cite{hub85} to measure similarity between community structures
generated by different algorithms as well as those from different runs of
the same algorithm. ARI scores close to $1$ are obtained when comparing the
results of multiple runs of the same algorithm on the LCC of the
co-inventor network, indicating the generally very good or, for Propagating
Labels, at least satisfactory robustness of each method. The ARI value of
$0.8$ found when comparing the community structures generated from the
Greedy and Louvain algorithms attests to their high degree of similarity.
Similarity to a much lesser extent is exhibited between the partitionings
arising from the Infomap and Propagating-Labels methods, as well as between
those of Greedy/Louvain and Random Walks. All other pairwise comparisons
yield a very small ARI score of $0.2$ and below.

Having identified inventor communities in the LCC on the co-inventor
network, we track the citations acquired by patents from the chosen cohort
over a ten-year period from each individual patent's time of grant.
Following the usual convention~\cite{hal02,meh10}, we assign the time of
citation to be the application date of the citing patent. Hence, citations
to patents in the cohort considered here can only originate from patents
applied for before 2010. All citation-related data are sourced directly
from the USPTO~\cite{uspto}. For the purpose of the present work, we
specifically consider the time $\Delta t_1$ elapsed after each patent's
time of grant until it acquires its first citation~\cite{gay05,lee17} and
determine this citation to be either a self-citation, an in-community
citation, or an out-of-community citation based on the previously obtained
community structure. Here a \textit{self}-citation occurs if any of the
inventors listed on the citing patent is also listed as an inventor on the
cited patent, which is the case independently of any community structure.
An \textit{in-community\/} citation is a \textit{non-self}-citation from a
patent where at least one of the listed inventors shares membership in one
of the algorithmically identified communities with at least one of the
inventors listed on the cited patent. An \textit{out-of-community\/}
citation is from a patent whose listed inventors belong to different
communities than the inventors of the cited patent. To focus most precisely
on how community structure affects the knowledge flow through the
co-inventor network, we only consider citations from patents that have at
least one of their listed inventors belonging to the LCC of the co-inventor
network defined via the cohort of cited patents\footnote{Excluding
citations by inventors outside the LCC amounts to neglecting trivial
out-of-community citations that would likely influence the time-lag
distribution for this type of citation mostly by shifting weight to larger
$\Delta t_1$~\cite{sin05,sor06,bre09}. If at all relevant, this could only
further accentuate the differences between characteristic values (mean,
median, and mode) for the time lag to first citation found for in-community
and out-of-community citations.}. Table~\ref{tab:summStat} provides
citation-related summary statistics for the analyzed patent cohort.

Distributions of time lags $\Delta t_1$ to first citation obtained for each
subset of citation type (self, in-community, and out-of-community) are
juxtaposed in Fig.~\ref{fig:lagDist}. Their distinctive properties are
discussed in greater detail in the next Section. Note that, because the
time of citation is the \emph{application\/} date of the citing patent, and
the time lag $\Delta t_1$ is measured from the \emph{grant\/} date of the
cited patent, negative values of $\Delta t_1$ are possible and, in fact,
occur quite frequently. That patents have often already accrued citations
by the time of grant from other patents that were applied for before that
date but granted afterwards is a well-known feature of patent-citation
dynamics~\cite{meh10}.

\section*{Results and Discussion}

As is apparent from the examples shown in Fig.~\ref{fig:lagDist},
distributions of time lags to first citation are generally broad and
skewed. We find that log-normal distributions provide a successful fit to
their line shapes, except for the distinctive peak for the $\Delta t_1 = 0$
bin exhibited by the time-lag distribution of first citations that are
self-citations. Further research is needed to elucidate the origin of the
inflated probability for a zero time lag for self-citations; we can only
speculate at this point that it could be the result of larger firms'
patenting strategies.

The analysis of time-lag distributions for in-community vs.\
out-of-community first citations enables us to discuss the influence of
community structure on citation dynamics, for each set of communities
identified by the five different algorithms used in this work.
Specifically, we consider distributional averages (mean and median values),
as well as the most probable value (mode). Table~\ref{tab:commResults}
gives the results obtained for these by two different methods: direct
calculation using the empirical data for the time-lag distributions, and
the values derived from parameters of the fitted log-normal distributions.
For the mean and median values of time lags for in-community first
citations, there is excellent agreement, within uncertainties, between the
two approaches for all five community structures considered. Similarly good
agreement exists between the medians and modes of time lags for
out-of-community first citations. The disagreement between the modes
calculated directly and from the log-normal fit for in-community
first-citation time lags obtained based on the Infomap and Propagating
Labels algorithms is due to the slightly inflated probability for the
$\Delta t_1 = 0$ bin exhibited in these cases. See Figs.~\ref{fig:lagDist}
and \ref{fig:algoComp}. The magnitude of the inflated peak at $\Delta t_1
= 0$ for in-community first-citation time lags is generally comparable to
the size of statistical fluctuations and overall much smaller than that
observed in the time-lag distribution of first self-citations. The
consistently higher mean obtained for the out-of-community first-citation
time lag using the raw data as compared with that derived from the
log-normal-distribution fit parameters can be traced to the fact that the
fit systematically underestimates probabilities in the tail of the time-lag
distribution for this type of citation. Overall, the magnitude for the
means and medians found in this work for the first-citation time lag agrees
very well with that reported previously in the literature~\cite{tra90,
gay05,fis17,lee17}.

The specific community structure determined for the co-inventor network is
observed to depend sensitively on the employed algorithm. The Greedy and
Louvain algorithms yield a comparatively small number of larger
communities, whereas the Infomap and Propagating-Labels algorithms identify
mostly much smaller communities. The community structure obtained using
Random Walks lies somewhat inbetween these two extremes. See
Fig.~\ref{fig:commDist} and Table~\ref{tab:commStat}. Interestingly, the
distributional properties of the time lag for in-community first citations,
and even more so those pertaining to out-of-community first citations, are
found to be quite similar irrespective of the significant differences
between the underlying inventor-community structures.
Figure~\ref{fig:algoComp} shows the distributions of time lags for
in-community citations based on the community structures obtained from
running the Greedy, Louvain, Random-Walks, and Propagating-Labels
algorithms on the LCC of the co-inventor network. Corresponding results for
the community structure identified by the Infomap algorithm are given in
Fig.~\ref{fig:lagDist}. The distributions of the time lag for
out-of-community first citations obtained based on the communities
identified by the five different algorithms are visually barely
distinguishable and therefore not shown. This suggests that algorithms
producing larger communities have generally agglomerated parts of the
co-inventor network that are separate communities as far as citing behavior
is concerned. Hence, our study provides another real-world verification of
the resolution limit~\cite{for07} associated with algorithms based on
optimizing modularity.

Comparison of the median and mean values for the time lags to first
citation found for the in-community and out-of-community types,
respectively, shows systematic differences. See
Table~\ref{tab:commResults}. In particular, the median time lags to
in-community first citations are about 2-3~months shorter than the
corresponding medians for out-of-community first citations. The difference
between the means of the time lag for in-community and out-of-community
first citations is about 2-5 months. Given that we have analyzed time lags
to first citation for patents whose inventors are all connected by prior
co-authorship of patents, this indicates a strong influence of community
structure on patent-citation dynamics and the associated knowledge flow.
Further comparison with the mean, median, and mode values found for the
time lag to first citations that are self-citations is very instructive.
See Table~\ref{tab:selfResults}. The mean and median calculated using the
raw data, including the inflated peak at $\Delta t_1 =0$, are found to be
essentially the same as for in-community citations. Hence, on average,
in-community first citations and first-self-citations occur after similar
time periods that are, again on average, shorter by about 3 months than the
time lag for out-of-community first citations. However, the distributions
of time lags for the in-community and self types of first citations differ
significantly. This can be illustrated by analyzing the adjusted
distribution of time lags for self-citations where the inflated probability
at $\Delta t_1 =0$ has been replaced with an interpolated value, or with
the peak removed entirely. The mean, median and mode values found for this
modified distribution agree almost exactly with those found for
out-of-community citations. See again Table~\ref{tab:selfResults}. This
interesting combination of properties invites further investigation. A
tentative speculation could be made that both the out-of-community type of
first citation and the first-self-citations outside the $\Delta t_1 =
0$ peak originate largely from examiners, whereas the in-community type and
the self-citations from the inflated $\Delta t_1 = 0$ peak are generally
made by inventors themselves. As inventor and examiner citations have been
separately identified on USPTO patents granted since 2001, a follow-up
study utilizing a large-enough cohort of post-2001 patents with
sufficiently long time window for citation accrual should soon be possible.

We employed Welch's \textit{t}-test~\cite{wel47} to establish whether the
difference between the means of the citation lag for in-community and
out-of-community citations is statistically significant. Results are
summarized in Table~\ref{tab:commResults}. The \textit{t}-scores ranging
between $6.6$ and $8.7$ indicate that the differences in the means of
citation lags are statistically extremely significant. Although the
\textit{t}-test is expected to yield reliable scores for non-normal
distributions~\cite{lum02}, we also applied it to the, to a good
approximation normal, distribution of the log of the time lag to first
citation (after shifting the latter by a constant to eliminate negative
values). Also in this case, similarly large \textit{t}-scores reject the
possibility of the means being equal with high confidence.

The high \textit{t}-values generated in our application of the Welch test
are likely a consequence of the very large sample sizes ($1\, 415$ data
points from in-community citations, $12\, 272$ from out-of-community
citations) that increase sensitivity to small differences. To illustrate
the effect sample size has on the outcome of the \textit{t}-test, we
selected a random and independently chosen subset of 300 citations from
each of the in-community and out-of-community datasets and performed
Welch's \textit{t}-test on these small samples. This was repeated 500 times
to obtain averages of such \textit{t}-values, which are also listed in
Table~\ref{tab:commResults}. In the case of the Infomap-generated community
partition, the mean \textit{t}-value resulting from this procedure is
$2.9$, along with $83$\% of the small-sample simulations finding
differences between the distributions' means to be significant to the $5$\%
$\alpha$ level. This indicates that sample sizes of at least 300 first
citations are needed to detect with sufficient confidence the time delay
reported here between those originating from within and outside of
co-inventor communities.

As further proof of the inventor-community-related cause of acceleration
in patent-citation dynamics, we performed the following control experiment.
Starting with the Infomap-generated community partition of the LCC,
inventors were randomly re-assigned to communities within this given
structure. As a result, the high-level morphology (number of communities
and their individual sizes) of the partition was left unchanged, but the
groupings of inventors within each community became completely random, as
indicated by the ARI value of $0.0002$ obtained when comparing the initial
and randomized partitions. Repeating the analysis of first citations, we
observe that the total number of in-community citations occurring in the
randomized community structure has dropped precipitously to $129$, i.e.,
about $10\,$\% of the in-community citations present in the original
structure. Furthermore, the mean (median) time to a first in-community
citation in the randomized structure is determined to be $23.9\,$months
($18\,$months), which is not significantly earlier anymore than the
corresponding value of $24.4\,$months ($19\,$months) found for the
out-of-community first citations. The Welch's \textit{t}-test score of
$0.24$ also indicates that there is no significant difference between the
means of the distributions of times to first in-community and
out-of-community citations in the situation with randomized inventor
assignment to communities. The significantly reduced total number of
in-community citations, and the disappearing difference between mean times
to first citations that originate from within and outside of communities,
both convincingly indicate that the originally established inventor
communities are indeed the platforms for accelerated patent-citation
dynamics.

We close this section by discussing several confounding variables whose
influence may weaken our inference of a general non-trivial
community-structure effect on citation dynamics. \textit{Inventor team
size\/}: All other things being equal, inventors working in larger teams
will have a greater chance of acquiring in-community first citations to
their patents, which would also be likely to occur more quickly. However,
the scope for this trivial mechanism causing the observed effect is
severely limited by the fact that large inventor teams are actually quite
rare. See, e.g., data presented in Ref.~\cite{cre16}. The average community
size found in our work certainly exceeds the average inventor-team size
given for the period 1995-1999 in Fig.~4 of that article.
\textit{Firm-level associations\/}: Certain inventor communities may just
be a reflection of the inventors' employment at the same firm and, for
their case, intra-firm information channels acting in parallel to prior
co-invention activity may drive the observed acceleration in the
in-community citation dynamics. However, for large firms, especially those
with multiple geographically separated R{\&}D centers, this alternative
mechanism could be ineffective. Disentangling trivial firm-association
effects from knowledge flows established via real inventor collaboration
would be an interesting direction for future research.
\textit{Niche-technology associations\/}: The patent cohort studied in this
work relates to the broad and technologically crowded semiconductor
industry, and our detected communities may correlate with very specific
technology types. The fact that citations are likely to come first from
inventors in the same community could then arise simply because inventors
that are working in the same industrial niche are likely to build on each
others' technological advances before inventors working in technologically
more distant fields. To clarify this issue, the distributions of technology
specializations across the respective in-community and out-of-community
citing-patent cohorts would need to be studied in greater detail using a
suitable proxy measure for technological similarity that could be defined
either within~\cite{tra97} or beyond~\cite{kay14} existing classifications
schemes. 

\section*{Conclusions}

We have investigated community structure on the co-inventor network
associated with a particular cohort of USPTO patents where edge weights
reflect the frequency of inventors' collaborative patenting activity. Five
established community-detection algorithms (Greedy, Louvain, Infomap,
Random Walks, and Propagating Labels) were deployed to identify communities
on this network's largest connected component. The sizes and numbers of
communities found by the different algorithms varied, with some
similarities exhibited by algorithms using related methodology.
Table~\ref{tab:commStat} and Fig.~\ref{fig:commDist} provide details
enabling a quantitative comparison between the properties of the
algorithmically detected community structures.

To investigate the effect of inventor communities on patent-citation
dynamics, we analyzed the time lag to the first citation received by the
patents associated with inventors from the largest connected component of
the co-inventor network. Only citations originating from patents
co-authored by at least one of the inventors from the largest connected
component were counted for the present study. Three different types of
first citation were distinguished: self-citations, in-community
(non-self-)citations, and out-of-community citations. The distributions of
time lags for each type of first citation were observed to have distinctive
properties. Figure~\ref{fig:lagDist} shows results obtained based on the
community structure found by the Infomap algorithm. The mean, median, and
mode values for each type of distribution were determined to enable a
quantitative comparison of the speeds of knowledge flow within and between
different inventor communities. Results for all community structures
investigated in the present work are summarized in
Table~\ref{tab:commResults}. The median time delay to first citation for
the out-of-community type turns out to be typically $3\,$months longer than
for the in-community type. Self-citations and in-community-type citations
have approximately the same median time lag, even though the distributions
of time lags for these two types of first citation are markedly different.
The difference between the mean time lag observed for out-of-community and
in-community first citations is generally even larger than that found for
the corresponding median values. Although the communities identified by the
different detection algorithms utilized in this work differ in some detail,
the observed influence of community structure on patent-citation dynamics
was found to agree closely, even on a quantitative level. Thus our results
provide a rather general quantification for the accelerated knowledge flow
through inventor communities formed via collaborative patenting.
Furthermore, the observation that association with distinct inventor
communities based on previous co-authorship on patents results in faster
citation of an inventor's later patents by members of that community
provides strong further evidence in support of the fundamental importance
of such collaboration-based social connections~\cite{sin05,agr06,bre09,
bre16} that has not always been able to be clearly observed~\cite{sto02}.

Our focus on the time lag to patents' first citation was motivated by the
expectation that this quantity will likely be the best proxy measure for
a real difference in time scales for knowledge propagation within and
outside of co-inventor communities. It would be interesting to also compare
the mean time lags between later (i.e., second, third, $\dots n$th)
in-community and out-of-community citations. Such a study should be able to
observe how the in-community advantage for knowing earlier about an
invention diminishes over its repeated utilization.

The results of our work point the way to other interesting directions for
future research. For example, analyzing the characteristics of the
algorithmically identified inventor communities could yield useful
information regarding the structure of effective innovation teams,
extending previous work that focused only on direct collaborations between
inventors~\cite{cre16}. Studying the dependence of the observed
acceleration of information flow through inventor communities on the field
and type of inventions may yield a measure for the speed at which the
knowledge frontier moves in different parts of innovation space.
Community-detection methods could also be deployed to elucidate
relationships shaping innovation activity beyond the network of inventors,
e.g., on the level of firms~\cite{sch07} or other organizations. Thus
opportunities abound for the useful application of modern network-analysis
tools to innovation economics~\cite{ace16} and related social-science
studies~\cite{hid16}.

%%%%%%%%%%%%%%%%%%%%%%%%%%%%%%%%%%%%%%%%%%%%%%%%%%%%%%%%%%%%%%%%%%%%%%%%%%%
%%                                                                       %%
%% Backmatter begins here                                                %%
%%                                                                       %%
%%%%%%%%%%%%%%%%%%%%%%%%%%%%%%%%%%%%%%%%%%%%%%%%%%%%%%%%%%%%%%%%%%%%%%%%%%%

\begin{backmatter}

\section*{Availability of data and material}
The datasets supporting the conclusions of this article are available in
the Zenodo repository,
\href{http://doi.org/10.5281/zenodo.2574342}{http://doi.org/10.5281/zenodo.2574342}.

\section*{Competing interests}
The authors declare that they have no competing interests.

\section*{Funding}
The work of WD was supported by a Victoria University of Wellington Faculty
of Science Strategic Grant (GMS grant no.\ 217470). Funding from the New
Zealand Tertiary Education Commission's Centres of Research Excellence Fund
via Te P{\=u}naha Matatini is gratefully acknowledged by KWH, MG and UZ.

\section*{Authors' contributions}
KWH, MG and UZ designed the research. WD constructed the co-inventor
network and performed the data analysis, with assistance from KWH. All
authors contributed to analyzing the results. UZ wrote the paper with input
from all authors.

\section*{Acknowledgements}
The authors obtained useful insights into companies' patenting practices
from conversations with L. Colombo and F. Natali, discussed
community-detection algorithms with S. Marsland and D. O'Neale, and
received advice on statistical testing from I. Doonan.

% if your bibliography is in bibtex format, use those commands:
%\bibliographystyle{bmc-mathphys} % Style BST file (bmc-mathphys, vancouver, spbasic).
%\bibliography{inventcomm} % Bibliography file (usually '*.bib' )
% for author-year bibliography (bmc-mathphys or spbasic)
% a) write to bib file (bmc-mathphys only)
% @settings{label, options="nameyear"}

% or include bibliography directly:

\newcommand{\BMCxmlcomment}[1]{}

\BMCxmlcomment{

<refgrp>

<bibl id="B1">
  <title><p>The {NBER} Patent Citations Data File: Lessons, Insights and
  Methodological Tools</p></title>
  <aug>
    <au><snm>Hall</snm><fnm>BH</fnm></au>
    <au><snm>Jaffe</snm><fnm>AB</fnm></au>
    <au><snm>Trajtenberg</snm><fnm>M</fnm></au>
  </aug>
  <source>Patents, Citations, and Innovations: A Window on the Knowledge
  Economy</source>
  <publisher>Cambridge, MA: MIT Press</publisher>
  <editor>A. B. Jaffe and M. Trajtenberg</editor>
  <pubdate>2002</pubdate>
  <fpage>403</fpage>
  <lpage>460</lpage>
</bibl>

<bibl id="B2">
  <title><p>Knowledge networks from patent data: Methodological issues and
  research targets</p></title>
  <aug>
    <au><snm>Breschi</snm><fnm>S</fnm></au>
    <au><snm>Lissoni</snm><fnm>F</fnm></au>
  </aug>
  <source>Handbook of Quantitative Science and Technology Research</source>
  <publisher>Dordrecht: Kluwer</publisher>
  <editor>Moed, Henk F. and Gl{\"a}nzel, Wolfgang and Schmoch, Ulrich</editor>
  <pubdate>2004</pubdate>
  <fpage>613</fpage>
  <lpage>643</lpage>
</bibl>

<bibl id="B3">
  <title><p>Patent citation data in social science research: {O}verview and
  best practices</p></title>
  <aug>
    <au><snm>Jaffe</snm><fnm>AB</fnm></au>
    <au><snm>{de Rassenfosse}</snm><fnm>G</fnm></au>
  </aug>
  <source>J. Assoc. Inf. Sci. Technol.</source>
  <pubdate>2017</pubdate>
  <volume>68</volume>
  <fpage>1360</fpage>
  <lpage>1374</lpage>
</bibl>

<bibl id="B4">
  <title><p>Determinants of knowledge diffusion as evidenced in patent data:
  the case of liquid crystal display technology</p></title>
  <aug>
    <au><snm>Stolpe</snm><fnm>M</fnm></au>
  </aug>
  <source>Res. Policy</source>
  <pubdate>2002</pubdate>
  <volume>31</volume>
  <fpage>1181</fpage>
  <lpage>1198</lpage>
</bibl>

<bibl id="B5">
  <title><p>Networks of inventors and the role of academia: an exploration of
  {I}talian patent data</p></title>
  <aug>
    <au><snm>Balconi</snm><fnm>M</fnm></au>
    <au><snm>Breschi</snm><fnm>S</fnm></au>
    <au><snm>Lissoni</snm><fnm>F</fnm></au>
  </aug>
  <source>Res. Policy</source>
  <pubdate>2004</pubdate>
  <volume>33</volume>
  <fpage>127</fpage>
  <lpage>145</lpage>
</bibl>

<bibl id="B6">
  <title><p>Collaborative Networks as Determinants of Knowledge Diffusion
  Patterns</p></title>
  <aug>
    <au><snm>Singh</snm><fnm>J</fnm></au>
  </aug>
  <source>Manage. Sci.</source>
  <pubdate>2005</pubdate>
  <volume>51</volume>
  <fpage>756</fpage>
  <lpage>770</lpage>
</bibl>

<bibl id="B7">
  <title><p>Complexity, networks and knowledge flow</p></title>
  <aug>
    <au><snm>Sorenson</snm><fnm>O</fnm></au>
    <au><snm>Rivkin</snm><fnm>JW</fnm></au>
    <au><snm>Fleming</snm><fnm>L</fnm></au>
  </aug>
  <source>Res. Policy</source>
  <pubdate>2006</pubdate>
  <volume>35</volume>
  <fpage>994</fpage>
  <lpage>1017</lpage>
</bibl>

<bibl id="B8">
  <title><p>Mobility of skilled workers and co-invention networks: an anatomy
  of localized knowledge flows</p></title>
  <aug>
    <au><snm>Breschi</snm><fnm>S</fnm></au>
    <au><snm>Lissoni</snm><fnm>F</fnm></au>
  </aug>
  <source>J. Econ. Geogr.</source>
  <pubdate>2009</pubdate>
  <volume>9</volume>
  <fpage>439</fpage>
  <lpage>468</lpage>
</bibl>

<bibl id="B9">
  <title><p>Co-invention networks and inventive productivity in {US}
  cities</p></title>
  <aug>
    <au><snm>Breschi</snm><fnm>S</fnm></au>
    <au><snm>Lenzi</snm><fnm>C</fnm></au>
  </aug>
  <source>J. Urban Econ.</source>
  <pubdate>2016</pubdate>
  <volume>92</volume>
  <fpage>66</fpage>
  <lpage>75</lpage>
</bibl>

<bibl id="B10">
  <title><p>Inventor collaboration and its persistence across {E}uropean
  regions</p></title>
  <aug>
    <au><snm>T{\'o}th</snm><fnm>G.</fnm></au>
    <au><snm>Juh{\'a}sz</snm><fnm>S.</fnm></au>
    <au><snm>Elekes</snm><fnm>Z.</fnm></au>
    <au><snm>Lengyel</snm><fnm>B.</fnm></au>
  </aug>
  <pubdate>2018</pubdate>
  <url>https://arxiv.org/abs/1807.07637</url>
  <note>{p}reprint arXiv:1807.07637</note>
</bibl>

<bibl id="B11">
  <title><p>The determinants of patent citations: an empirical analysis of
  {F}rench and {B}ritish patents in the {US}</p></title>
  <aug>
    <au><snm>Gay</snm><fnm>C.</fnm></au>
    <au><snm>{Le Bas}</snm><fnm>C.</fnm></au>
    <au><snm>Patel</snm><fnm>P.</fnm></au>
    <au><snm>Touach</snm><fnm>K.</fnm></au>
  </aug>
  <source>Econ. Innov. New Technol.</source>
  <pubdate>2005</pubdate>
  <volume>14</volume>
  <fpage>339</fpage>
  <lpage>350</lpage>
</bibl>

<bibl id="B12">
  <title><p>What makes the first forward citation of a patent occur
  earlier?</p></title>
  <aug>
    <au><snm>Lee</snm><fnm>J</fnm></au>
    <au><snm>Sohn</snm><fnm>SY</fnm></au>
  </aug>
  <source>Scientometrics</source>
  <pubdate>2017</pubdate>
  <volume>113</volume>
  <fpage>279</fpage>
  <lpage>298</lpage>
</bibl>

<bibl id="B13">
  <title><p>Geographic Localization of Knowledge Spillovers as Evidenced by
  Patent Citations</p></title>
  <aug>
    <au><snm>Jaffe</snm><fnm>AB</fnm></au>
    <au><snm>Trajtenberg</snm><fnm>M</fnm></au>
    <au><snm>Henderson</snm><fnm>R</fnm></au>
  </aug>
  <source>Quart. J. Econ.</source>
  <pubdate>1993</pubdate>
  <volume>108</volume>
  <fpage>577</fpage>
  <lpage>598</lpage>
</bibl>

<bibl id="B14">
  <title><p>Knowledge of the Firm, Combinative Capabilities, and the
  Replication of Technology</p></title>
  <aug>
    <au><snm>Kogut</snm><fnm>B</fnm></au>
    <au><snm>Zander</snm><fnm>U</fnm></au>
  </aug>
  <source>Organ. Sci.</source>
  <pubdate>1992</pubdate>
  <volume>3</volume>
  <fpage>383</fpage>
  <lpage>397</lpage>
</bibl>

<bibl id="B15">
  <title><p>Community detection in graphs</p></title>
  <aug>
    <au><snm>Fortunato</snm><fnm>S</fnm></au>
  </aug>
  <source>Phys. Rep.</source>
  <pubdate>2010</pubdate>
  <volume>486</volume>
  <fpage>75</fpage>
  <lpage>174</lpage>
</bibl>

<bibl id="B16">
  <title><p>Community detection in networks: {A} user guide</p></title>
  <aug>
    <au><snm>Fortunato</snm><fnm>S</fnm></au>
    <au><snm>Hric</snm><fnm>D</fnm></au>
  </aug>
  <source>Phys. Rep.</source>
  <pubdate>2016</pubdate>
  <volume>659</volume>
  <fpage>1</fpage>
  <lpage>44</lpage>
</bibl>

<bibl id="B17">
  <title><p>Communities, knowledge creation, and information
  diffusion</p></title>
  <aug>
    <au><snm>Lambiotte</snm><fnm>R.</fnm></au>
    <au><snm>Panzarasa</snm><fnm>P.</fnm></au>
  </aug>
  <source>J. Informetrics</source>
  <pubdate>2009</pubdate>
  <volume>3</volume>
  <fpage>180</fpage>
  <lpage>190</lpage>
</bibl>

<bibl id="B18">
  <title><p>Community structure of the physical review citation
  network</p></title>
  <aug>
    <au><snm>Chen</snm><fnm>P.</fnm></au>
    <au><snm>Redner</snm><fnm>S.</fnm></au>
  </aug>
  <source>J. Informetrics</source>
  <pubdate>2010</pubdate>
  <volume>4</volume>
  <fpage>278</fpage>
  <lpage>290</lpage>
</bibl>

<bibl id="B19">
  <title><p>Knowledge evolution in physics research: {A}n analysis of
  bibliographic coupling networks</p></title>
  <aug>
    <au><snm>Liu</snm><fnm>A</fnm></au>
    <au><snm>Cheong</snm><fnm>SA</fnm></au>
  </aug>
  <source>PLoS ONE</source>
  <pubdate>2017</pubdate>
  <volume>12</volume>
  <fpage>e0184821</fpage>
</bibl>

<bibl id="B20">
  <title><p>Community Detection and Growth Potential Prediction Using the
  {S}tochastic {B}lock {M}odel and the Long Short-term Memory from Patent
  Citation Networks</p></title>
  <aug>
    <au><snm>Nakai</snm><fnm>K.</fnm></au>
    <au><snm>Nonaka</snm><fnm>H.</fnm></au>
    <au><snm>Hentona</snm><fnm>A.</fnm></au>
    <au><snm>Kanai</snm><fnm>Y.</fnm></au>
    <au><snm>Sakumoto</snm><fnm>T.</fnm></au>
    <au><snm>Kataoka</snm><fnm>S.</fnm></au>
    <au><snm>Carreón</snm><fnm>E. C. A.</fnm></au>
    <au><snm>Hiraoka</snm><fnm>T.</fnm></au>
  </aug>
  <source>2018 IEEE International Conference on Industrial Engineering and
  Engineering Management (IEEM)</source>
  <pubdate>2018</pubdate>
  <fpage>1884</fpage>
  <lpage>1888</lpage>
</bibl>

<bibl id="B21">
  <title><p>Community evolution in patent networks: technological change and
  network dynamics</p></title>
  <aug>
    <au><snm>Gao</snm><fnm>Y</fnm></au>
    <au><snm>Zhu</snm><fnm>Z</fnm></au>
    <au><snm>Kali</snm><fnm>R</fnm></au>
    <au><snm>Riccaboni</snm><fnm>M</fnm></au>
  </aug>
  <source>Appl. Netw. Sci.</source>
  <pubdate>2018</pubdate>
  <volume>3</volume>
  <fpage>26</fpage>
</bibl>

<bibl id="B22">
  <title><p>Disambiguation and co-authorship networks of the {U.S.} patent
  inventor database (1975--2010)</p></title>
  <aug>
    <au><snm>Li</snm><fnm>GC</fnm></au>
    <au><snm>Lai</snm><fnm>R</fnm></au>
    <au><snm>{D'Amour}</snm><fnm>A</fnm></au>
    <au><snm>Doolin</snm><fnm>DM</fnm></au>
    <au><snm>Sun</snm><fnm>Y</fnm></au>
    <au><snm>Torvik</snm><fnm>VI</fnm></au>
    <au><snm>Yu</snm><fnm>AZ</fnm></au>
    <au><snm>Fleming</snm><fnm>L</fnm></au>
  </aug>
  <source>Res. Policy</source>
  <pubdate>2014</pubdate>
  <volume>43</volume>
  <fpage>941</fpage>
  <lpage>955</lpage>
</bibl>

<bibl id="B23">
  <title><p>{US} {P}atent {C}lassification</p></title>
  <source>\url{https://www.uspto.gov/web/patents/classification/selectnumwithtitle.htm}</source>
  <note>Accessed 8 November 2018.</note>
</bibl>

<bibl id="B24">
  <title><p>Scientific collaboration networks. {II. S}hortest paths, weighted
  networks, and centrality</p></title>
  <aug>
    <au><snm>Newman</snm><fnm>M. E. J.</fnm></au>
  </aug>
  <source>Phys. Rev. E</source>
  <pubdate>2001</pubdate>
  <volume>64</volume>
  <fpage>016132</fpage>
</bibl>

<bibl id="B25">
  <title><p>Scientific collaboration networks. {I. N}etwork construction and
  fundamental results</p></title>
  <aug>
    <au><snm>Newman</snm><fnm>M. E. J.</fnm></au>
  </aug>
  <source>Phys. Rev. E</source>
  <pubdate>2001</pubdate>
  <volume>64</volume>
  <fpage>016131</fpage>
</bibl>

<bibl id="B26">
  <title><p>Fast algorithm for detecting community structure in
  networks</p></title>
  <aug>
    <au><snm>Newman</snm><fnm>M. E. J.</fnm></au>
  </aug>
  <source>Phys. Rev. E</source>
  <pubdate>2004</pubdate>
  <volume>69</volume>
  <fpage>066133</fpage>
</bibl>

<bibl id="B27">
  <title><p>Fast unfolding of communities in large networks</p></title>
  <aug>
    <au><snm>Blondel</snm><fnm>VD</fnm></au>
    <au><snm>Guillaume</snm><fnm>{J.-L.}</fnm></au>
    <au><snm>Lambiotte</snm><fnm>R</fnm></au>
    <au><snm>Lefebvre</snm><fnm>E</fnm></au>
  </aug>
  <source>J. Stat. Mech.</source>
  <pubdate>2008</pubdate>
  <fpage>P10008</fpage>
</bibl>

<bibl id="B28">
  <title><p>Maps of random walks on complex networks reveal community
  structure</p></title>
  <aug>
    <au><snm>Rosvall</snm><fnm>M</fnm></au>
    <au><snm>Bergstrom</snm><fnm>CT</fnm></au>
  </aug>
  <source>Proc. Natl. Acad. Sci. U.S.A.</source>
  <pubdate>2008</pubdate>
  <volume>105</volume>
  <fpage>1118</fpage>
  <lpage>1123</lpage>
</bibl>

<bibl id="B29">
  <title><p>Computing Communities in Large Networks Using Random
  Walks</p></title>
  <aug>
    <au><snm>Pons</snm><fnm>P</fnm></au>
    <au><snm>Latapy</snm><fnm>M</fnm></au>
  </aug>
  <source>J. Graph Algorithms Appl.</source>
  <pubdate>2006</pubdate>
  <volume>10</volume>
  <fpage>191</fpage>
  <lpage>218</lpage>
</bibl>

<bibl id="B30">
  <title><p>Near linear time algorithm to detect community structures in
  large-scale networks</p></title>
  <aug>
    <au><snm>Raghavan</snm><fnm>UN</fnm></au>
    <au><snm>Albert</snm><fnm>R</fnm></au>
    <au><snm>Kumara</snm><fnm>S</fnm></au>
  </aug>
  <source>Phys. Rev. E</source>
  <pubdate>2007</pubdate>
  <volume>76</volume>
  <fpage>036106</fpage>
</bibl>

<bibl id="B31">
  <title><p>Resolution limit in community detection</p></title>
  <aug>
    <au><snm>Fortunato</snm><fnm>S</fnm></au>
    <au><snm>Barth{\'e}lemy</snm><fnm>M</fnm></au>
  </aug>
  <source>Proc. Natl. Acad. Sci. U.S.A.</source>
  <pubdate>2007</pubdate>
  <volume>104</volume>
  <fpage>36</fpage>
  <lpage>41</lpage>
</bibl>

<bibl id="B32">
  <title><p>Finding and evaluating community structure in networks</p></title>
  <aug>
    <au><snm>Newman</snm><fnm>M. E. J.</fnm></au>
    <au><snm>Girvan</snm><fnm>M.</fnm></au>
  </aug>
  <source>Phys. Rev. E</source>
  <pubdate>2004</pubdate>
  <volume>69</volume>
  <fpage>026113</fpage>
</bibl>

<bibl id="B33">
  <title><p>Comparing Partitions</p></title>
  <aug>
    <au><snm>Hubert</snm><fnm>L.</fnm></au>
    <au><snm>Arabie</snm><fnm>P.</fnm></au>
  </aug>
  <source>J. Classification</source>
  <pubdate>1985</pubdate>
  <volume>2</volume>
  <fpage>193</fpage>
  <lpage>218</lpage>
</bibl>

<bibl id="B34">
  <title><p>Identifying the age profile of patent citations: new estimates of
  knowledge diffusion</p></title>
  <aug>
    <au><snm>Mehta</snm><fnm>A</fnm></au>
    <au><snm>Rysman</snm><fnm>M</fnm></au>
    <au><snm>Simcoe</snm><fnm>T</fnm></au>
  </aug>
  <source>J. Appl. Econometrics</source>
  <pubdate>2010</pubdate>
  <volume>25</volume>
  <fpage>1179</fpage>
  <lpage>1204</lpage>
</bibl>

<bibl id="B35">
  <title><p>{USPTO Bulk Data Storage System}</p></title>
  <source>\url{https://bulkdata.uspto.gov/}</source>
  <note>Accessed 8 November 2018.</note>
</bibl>

<bibl id="B36">
  <title><p>A penny for your quotes: patent citations and the value of
  innovations</p></title>
  <aug>
    <au><snm>Trajtenberg</snm><fnm>M</fnm></au>
  </aug>
  <source>RAND J. Econom.</source>
  <pubdate>1990</pubdate>
  <volume>21</volume>
  <fpage>172</fpage>
  <lpage>187</lpage>
</bibl>

<bibl id="B37">
  <title><p>The value of {C}hinese patents: An empirical investigation of
  citation lags</p></title>
  <aug>
    <au><snm>Fisch</snm><fnm>C</fnm></au>
    <au><snm>Sandner</snm><fnm>P</fnm></au>
    <au><snm>Regner</snm><fnm>L</fnm></au>
  </aug>
  <source>China Econ. Rev.</source>
  <pubdate>2017</pubdate>
  <volume>45</volume>
</bibl>

<bibl id="B38">
  <title><p>The generalization of `{S}tudent's' problem When several different
  population variances are involved</p></title>
  <aug>
    <au><snm>Welch</snm><fnm>B. L.</fnm></au>
  </aug>
  <source>Biometrika</source>
  <pubdate>1947</pubdate>
  <volume>34</volume>
  <fpage>28</fpage>
  <lpage>35</lpage>
</bibl>

<bibl id="B39">
  <title><p>The Importance of the Normality Assumption in Large Public Health
  Data Sets</p></title>
  <aug>
    <au><snm>Lumley</snm><fnm>T</fnm></au>
    <au><snm>Diehr</snm><fnm>P</fnm></au>
    <au><snm>Emerson</snm><fnm>S</fnm></au>
    <au><snm>Chen</snm><fnm>L</fnm></au>
  </aug>
  <source>Annu. Rev. Public Health</source>
  <pubdate>2002</pubdate>
  <volume>23</volume>
  <fpage>151</fpage>
  <lpage>169</lpage>
</bibl>

<bibl id="B40">
  <title><p>Do inventors talk to strangers? {On} proximity and collaborative
  knowledge creation</p></title>
  <aug>
    <au><snm>Crescenzi</snm><fnm>R</fnm></au>
    <au><snm>Nathan</snm><fnm>M</fnm></au>
    <au><snm>Rodr{\'i}guez Pose</snm><fnm>A</fnm></au>
  </aug>
  <source>Res. Policy</source>
  <pubdate>2016</pubdate>
  <volume>45</volume>
  <fpage>177</fpage>
  <lpage>194</lpage>
</bibl>

<bibl id="B41">
  <title><p>University versus corporate patents: {A} window on the basicness of
  invention</p></title>
  <aug>
    <au><snm>Trajtenberg</snm><fnm>M</fnm></au>
    <au><snm>Henderson</snm><fnm>R</fnm></au>
    <au><snm>Jaffe</snm><fnm>A</fnm></au>
  </aug>
  <source>Econ. Innov. New Technol.</source>
  <pubdate>1997</pubdate>
  <volume>5</volume>
  <fpage>19</fpage>
  <lpage>50</lpage>
</bibl>

<bibl id="B42">
  <title><p>Patent overlay mapping: {V}isualizing technological
  distance</p></title>
  <aug>
    <au><snm>Kay</snm><fnm>L</fnm></au>
    <au><snm>Newman</snm><fnm>N</fnm></au>
    <au><snm>Youtie</snm><fnm>J</fnm></au>
    <au><snm>Porter</snm><fnm>AL</fnm></au>
    <au><snm>Rafols</snm><fnm>I</fnm></au>
  </aug>
  <source>J. Assoc. Inf. Sci. Technol.</source>
  <pubdate>2014</pubdate>
  <volume>65</volume>
  <fpage>2432</fpage>
  <lpage>2443</lpage>
</bibl>

<bibl id="B43">
  <title><p>Gone but not forgotten: knowledge flows, labor mobility, and
  enduring social relationships</p></title>
  <aug>
    <au><snm>Agrawal</snm><fnm>A</fnm></au>
    <au><snm>Cockburn</snm><fnm>I</fnm></au>
    <au><snm>McHale</snm><fnm>J</fnm></au>
  </aug>
  <source>J. Econ. Geogr.</source>
  <pubdate>2006</pubdate>
  <volume>6</volume>
  <fpage>571</fpage>
  <lpage>591</lpage>
</bibl>

<bibl id="B44">
  <title><p>Interfirm Collaboration Networks: The Impact of Large-Scale Network
  Structure on Firm Innovation</p></title>
  <aug>
    <au><snm>Schilling</snm><fnm>MA</fnm></au>
    <au><snm>Phelps</snm><fnm>CC</fnm></au>
  </aug>
  <source>Manage. Sci.</source>
  <pubdate>2007</pubdate>
  <volume>53</volume>
  <fpage>1113</fpage>
  <lpage>1126</lpage>
</bibl>

<bibl id="B45">
  <title><p>Innovation network</p></title>
  <aug>
    <au><snm>Acemoglu</snm><fnm>D</fnm></au>
    <au><snm>Akcigit</snm><fnm>U</fnm></au>
    <au><snm>Kerr</snm><fnm>WR</fnm></au>
  </aug>
  <source>Proc. Natl. Acad. Sci. U.S.A.</source>
  <pubdate>2016</pubdate>
  <volume>113</volume>
  <fpage>11483</fpage>
  <lpage>11488</lpage>
</bibl>

<bibl id="B46">
  <title><p>Disconnected, fragmented, or united? a trans-disciplinary review of
  network science</p></title>
  <aug>
    <au><snm>Hidalgo</snm><fnm>CA</fnm></au>
  </aug>
  <source>Appl. Netw. Sci.</source>
  <pubdate>2016</pubdate>
  <volume>1</volume>
  <fpage>6</fpage>
</bibl>

<bibl id="B47">
  <title><p>Gephi: An Open Source Software for Exploring and Manipulating
  Networks</p></title>
  <aug>
    <au><snm>Bastian</snm><fnm>M</fnm></au>
    <au><snm>Heymann</snm><fnm>S</fnm></au>
    <au><snm>Jacomy</snm><fnm>M</fnm></au>
  </aug>
  <source>Proceedings of the Third International AAAI Conference on Weblogs and
  Social Media</source>
  <publisher>AAAI Press</publisher>
  <editor>Eytan Adar and Matthew Hurst and Tim Finin and Natalie Glance and
  Nicolas Nicolov and Belle Tseng</editor>
  <fpage>361</fpage>
  <lpage>362</lpage>
</bibl>

</refgrp>
} % end of \BMCxmlcomment

\section*{Figures}
\begin{figure}[h!]
\includegraphics[width=0.75\textwidth]{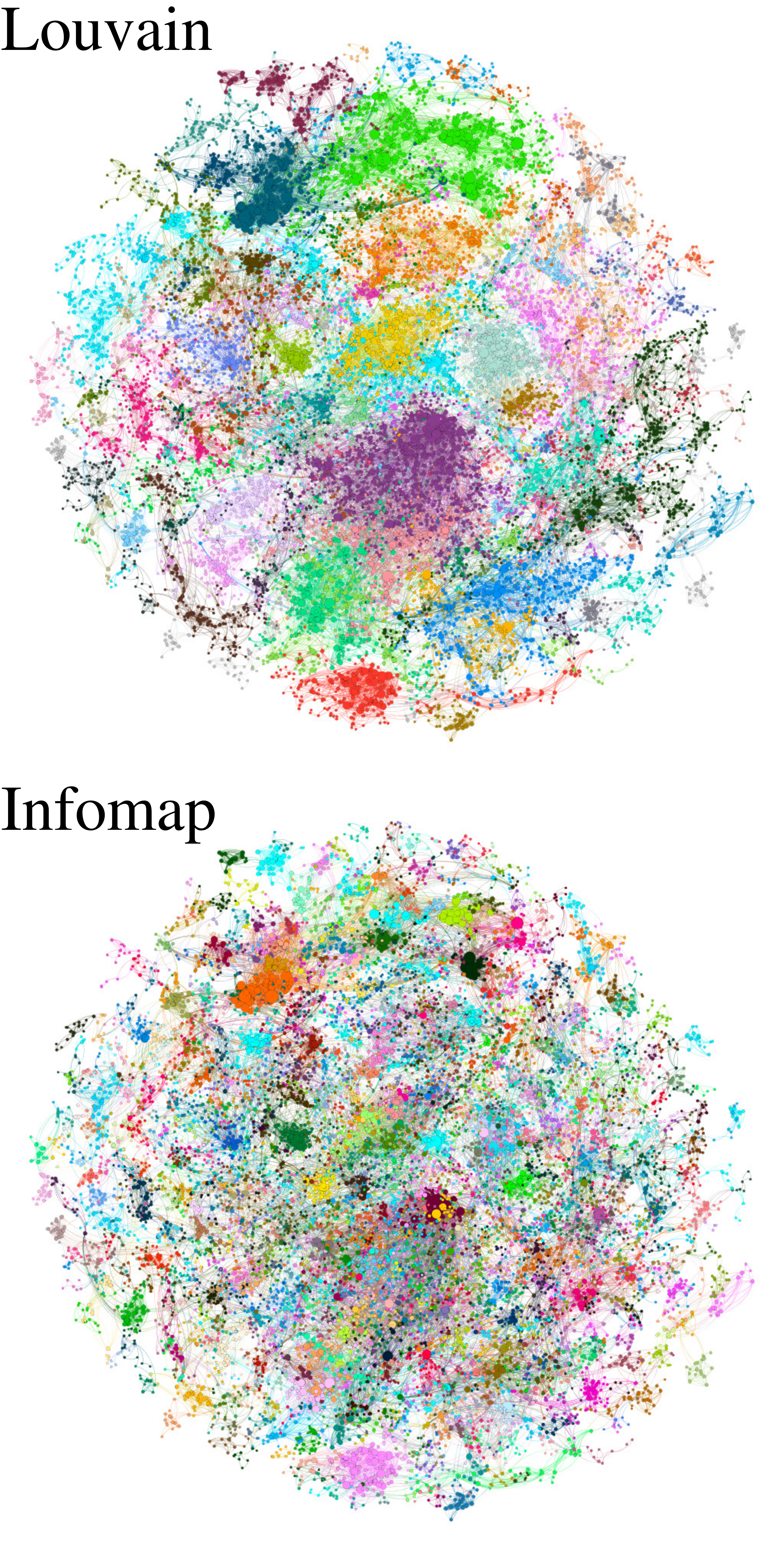}
\caption{\csentence{Community structure within the co-inventor network.}
The upper (lower) panel indicates communities in the largest connected
component of the co-inventor network identified by the Louvain (Infomap)
algorithm using different colors (but with some colors being reused in the
lower panel due to the overall large number of communities yielded by
Infomap). Note how certain groups of communities identified as separate by
Infomap are clustered together by Louvain. The size of the circle
indicating a node is proportional to the number of edges attached to that
node. Images created using Gephi~\cite{bas09}.%
\label{fig:LCCplots}}
\end{figure}

\begin{figure}[h!]
\includegraphics[width=0.95\textwidth]{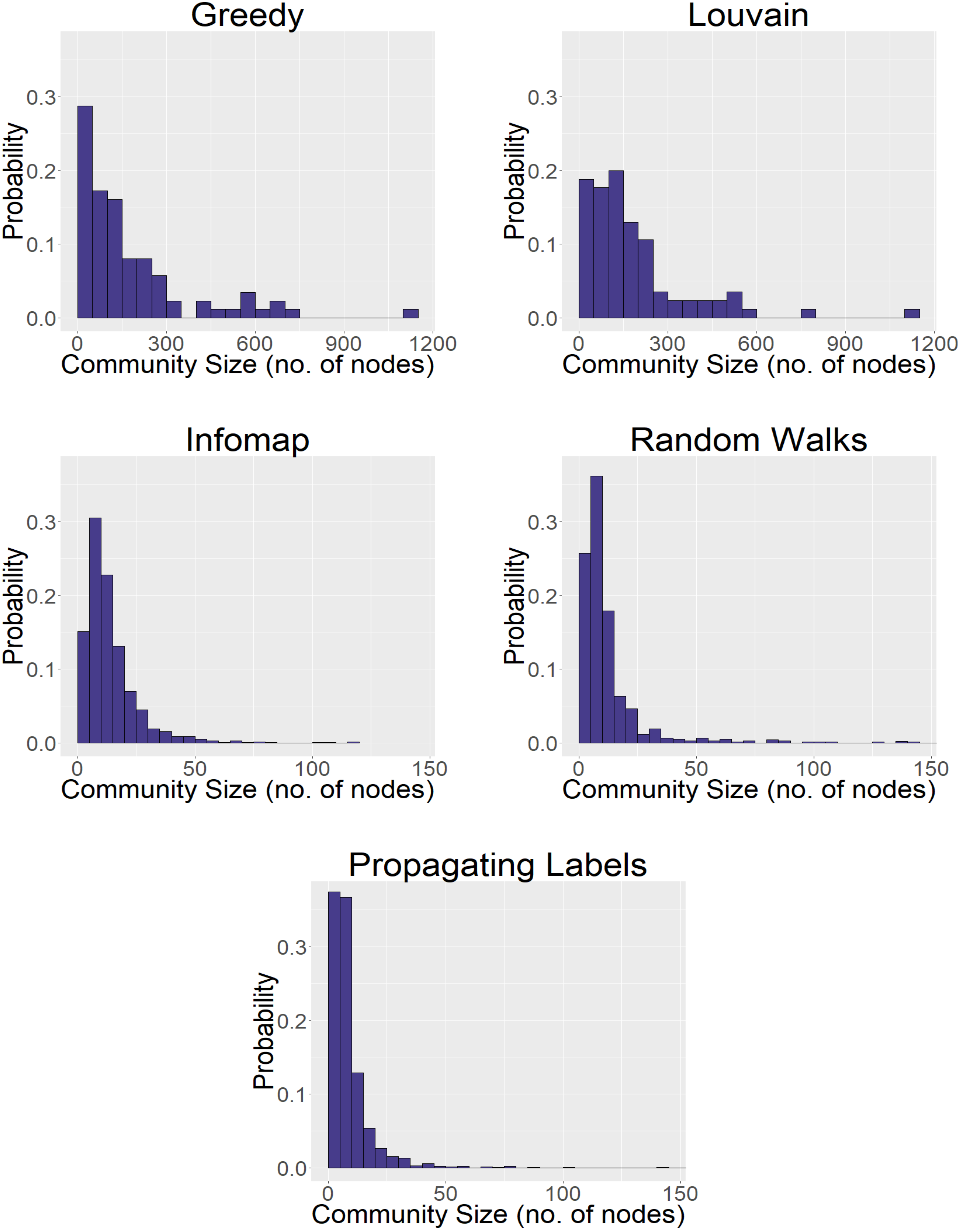}
\caption{\csentence{Size distributions of inventor communities found by
different algorithms.}
Distributions for the number of nodes within the communities identified on
the largest connected component of the co-inventor network by the Greedy,
Louvain, Infomap, Random-Walks, and Propagating-Labels algorithms. Note the
different bin widths: 50 (5) nodes in the plots in the first (second,
third) row. For Propagating Labels (Random Walks), one community with $262$
nodes (twelve communities with $163$, $169$, $172$, $175$, $194$, $222$,
$346$, $387$, $393$, $583$, $656$, and $1\, 060$ nodes, respectively) is
(are) outside the displayed range. See also Table~\ref{tab:commStat} for
community-structure-related summary statistics.%
\label{fig:commDist}}
\end{figure}

\begin{figure}[h!]
\includegraphics[width=0.95\textwidth]{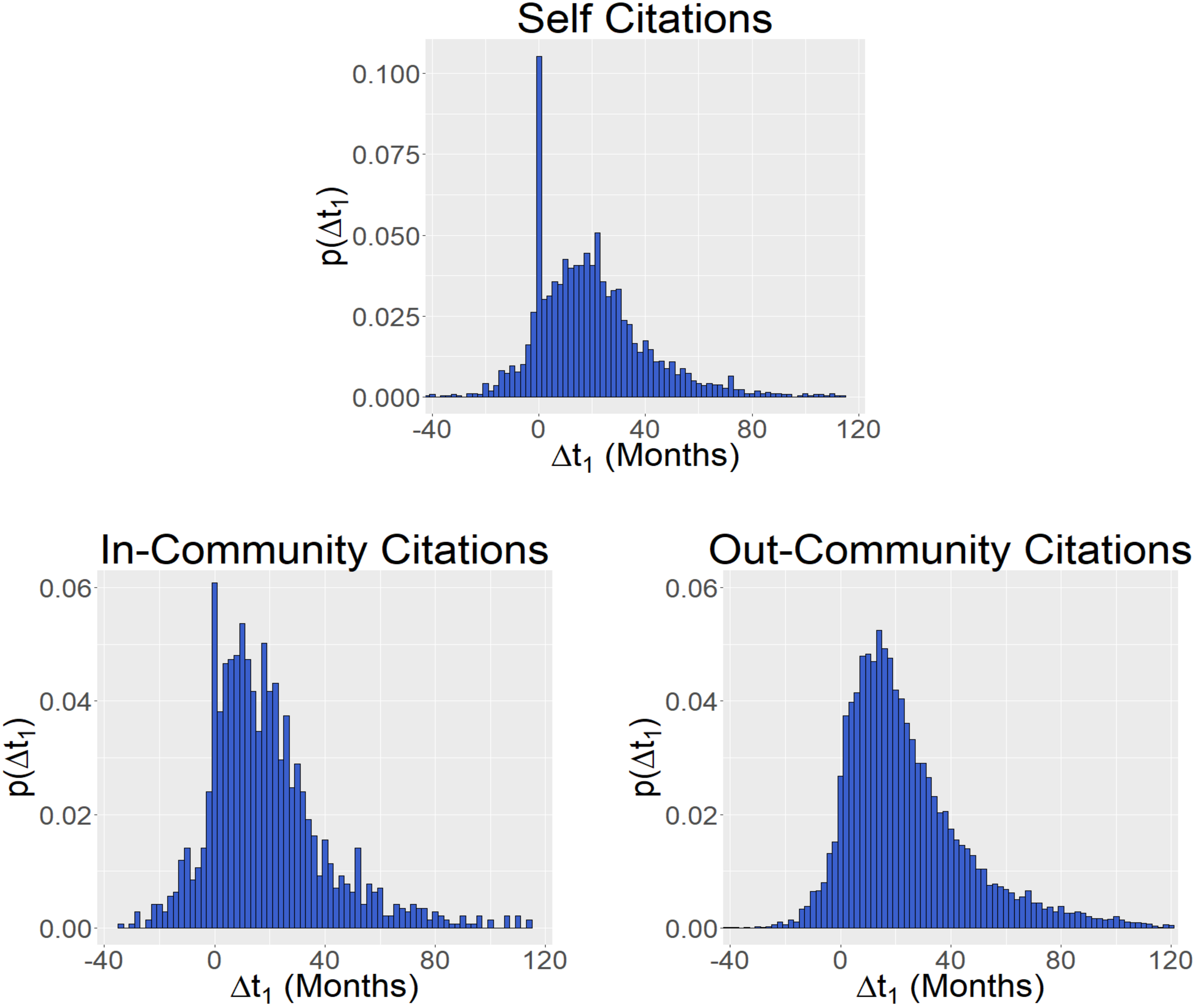}
\caption{\csentence{Time to first citation differentiated according to
origin of the first citation.}
The empirically obtained distributions of the time lag $\Delta t_1$ until
first citation are shown for the three different categories of first
citations considered in this work: self-citations, in-community citations,
and out-of-community citations. Results shown here for the latter two types
of citation have been obtained based on the community structure identified
by the Infomap algorithm on the largest connected component of the
co-inventor network. The bin width is 2 months.%
\label{fig:lagDist}}
\end{figure}

\begin{figure}[h!]
\includegraphics[width=0.95\textwidth]{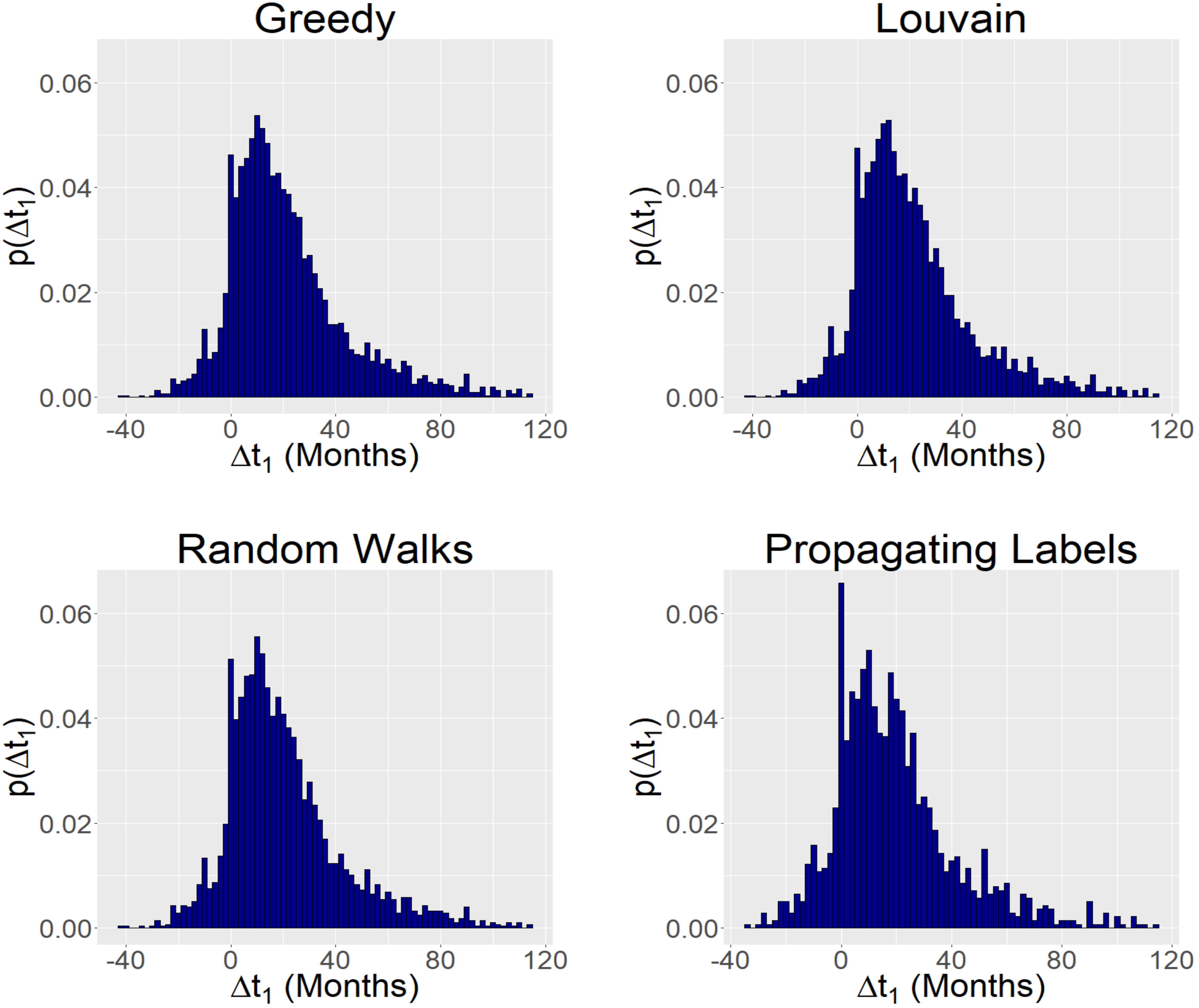}
\caption{\csentence{In-community first-citation time lag distributions for
different community structures.}
The distributions of the time lag $\Delta t_1$ until first citation when
this is an in-community citation are shown for the different community
structures on the largest connected component of the co-inventor network
obtained by the Greedy, Louvain Random-Walks, and Propagating-Labels
methods.%
\label{fig:algoComp}}
\end{figure}

\section*{Tables}
\begin{table}[h!]
\renewcommand*{\arraystretch}{1.5}
\caption{\label{tab:abbrevs}%
List of abbreviations.}
\begin{tabular}{|c|l|} \hline
abbreviation & meaning \\ \hline
ARI & adjusted Rand index \\
LCC & largest connected component \\
USPC & US Patent Classification \\
USPTO & US Patent and Trademark Office \\ \hline \end{tabular}
\end{table}

\begin{table}[h!]
\renewcommand*{\arraystretch}{1.5}
\caption{\label{tab:summStat}%
Summary statistics of patent data utilized for this study. The full patent
cohort comprises the USPTO patents from USPC classes 257, 326, and 438
granted during 1995-1999. Citations are tracked for each patent during the
10-year period after its time of grant.}
\begin{tabular}{|l|c|c|} \hline
{} & no.\ of patents & no.\ of inventors \\ \hline
full patent cohort & $46\, 034$ & $49\, 703$ \\
patents gaining a first citation within 10 years &  $44\,137$ & ---  \\
largest connected component (LCC) of co-inventor network & --- & $15\, 525$
\\ patents associated with inventors from the LCC & $20\, 021$ & --- \\
LCC-associated patents gaining a first citation within 10 years &
$19\, 284$ & --- \\
LCC-associated patents with a first citation by LCC inventors & $16\, 271$
& --- \\
Subset of these patents whose first citation is a self-citation & $2\, 584$
& --- \\ \hline \end{tabular}
\end{table}

\begin{table}[h!]
\renewcommand*{\arraystretch}{1.5}
\caption{\label{tab:commStat}%
Comparison of inventor-community structures found by different algorithms.
Summary statistics provided here pertain to the communities obtained by
applying each of the five indicated community-detection algorithms to the
largest connected component of the co-inventor network considered in this
work. Similarity between community structures is quantified in terms of the
adjusted Rand index (ARI)~\cite{hub85}. Listed ARI values are averages
calculated for pairs of community structures generated by multiple runs of
the respective algorithms that are being compared. The total number of
first citations classified as in-community citations based on each of the
five community partitions is also given.}
\begin{tabular}{|l|c|c|c|c|c|} \hline
{} & Greedy & Louvain & Infomap & Rand.\ Wlks.\ & Prop.\ Lbls.\ \\ \hline
Total no.\ of communities identified & $87$ & $85$ & $1\, 157$ & $948$ &
$1\, 800$ \\
Size of largest community (nodes) & $1\, 143$ & $1\, 139$ & $120$ &
$1\, 060$ & $262$ \\
simliarity to Greedy (ARI) & $1.0$ & $0.8$ & $0.1$ & $0.5$ & $0.1$ \\
similarity to Louvain (ARI) & $0.8$ & $1.0$ & $0.1$ & $0.5$ & $0.1$ \\
similarity to Infomap (ARI) & $0.1$ & $0.1$ & $0.9$ & $0.2$ & $0.6$ \\
similarity to Rand.\ Wlks.\ (ARI) & $0.5$ & $0.5$ & $0.2$ & $1.0$ & $0.2$
\\
similarity to Prop.\ Lbls.\ (ARI) & $0.1$ & $0.1$ & $0.6$ & $0.2$ & $0.7$
\\
no.\ of in-community first citations & $3\, 179$  & $3\, 031$ & $1\, 415$
& $2\, 772$ & $1\, 398$ \\ \hline
\end{tabular}
\end{table}

\begin{table}[h!]
\renewcommand*{\arraystretch}{1.5}
\caption{\label{tab:commResults}%
Community-related differences in the time lag $\Delta t_1$ until first
citation. The mean ($\overline{\Delta t_1}$), median ($\widetilde{\Delta
t_1}$), and mode ($\mathrm{Mo}_{\Delta t_1}$) values derived from the raw
data for the empirically observed distributions of time until the first
citation that is not a self-citation and either originates from within (In)
or outside (Out) co-inventor communities are listed for community
structures obtained using five different detection algorithms.
Corresponding values derived from fits of the empirical distributions to
log-normal form are also given for comparison. Welch's \textit{t}-test
scores~\cite{wel47} establish the statistical significance for the observed
differences between mean time lags for in-community and out-of-community
citations. We give \textit{t}-values obtained from tests applied to the
full datasets of in-community and out-of-community citation lags, as well
as the average \textit{t}-score found when comparing 500 independently and
randomly chosen subsets of 300 citations from the in- and out-of-community
datasets.}
\begin{minipage}{\textwidth}
\begin{tabular}{|l|l|c|c|c|c|c|}\hline
\multicolumn{2}{|c|}{} & Infomap & Prop.\ Lbls.\ & Rand.\ Wlks.\  & Greedy
& Louvain \\ \hline
\multirow{6}{2em}{In} & raw data $\overline{\Delta t_1}$ & $19.4$ & $19.7$
& $20.8$ & $21.6$ & $21.7$ \\
& raw data $\widetilde{\Delta t_1}$ & $16$ & $16$ & $17$ & $17$ & $17$ \\
& raw data\footnote{binned into 2-months-wide bins as shown in
  Figs.~\ref{fig:lagDist} and \ref{fig:algoComp}} $\mathrm{Mo}_{\Delta
  t_1}$ & $0$ & $0$ & $10$ & $10$& $12$ \\
& lognorm fit $\overline{\Delta t_1}$ & $18\pm2$ & $18\pm2$ & $18.8\pm0.9$
& $19.5\pm0.9$ & $19.5\pm0.9$ \\
& lognorm fit $\widetilde{\Delta t_1}$ & $16\pm1$ & $16\pm1$ & $16.7\pm0.7$
& $17.4\pm0.7$ & $17.4\pm0.7$ \\
& lognorm fit $\mathrm{Mo}_{\Delta t_1}$ & $12\pm2$ & $12\pm2$ &
$12.6\pm0.9$ & $12\pm2$ & $13.3\pm0.9$ \\\hline
\multirow{6}{2em}{Out} & raw data $\overline{\Delta t_1}$ & $24.6$ & $24.5$
& $24.8$ & $24.7$& $24.7$ \\
& raw data $\widetilde{\Delta t_1}$ & $19$ & $19$ & $20$ & $20$ & $20$ \\
& raw data $\mathrm{Mo}_{\Delta t_1}$& $15$& $15$ & $15$ & $15$ & $15$ \\
& lognorm fit $\overline{\Delta t_1}$ & $21.5\pm0.9$ & $21.5\pm0.9$ &
  $21.5\pm0.9$ & $21.5\pm0.9$ & $21.5\pm0.9$ \\
& lognorm fit $\widetilde{\Delta t_1}$ & $19.4\pm0.7$ & $19.4\pm0.7$ &
  $19.4\pm0.7$ & $19.4\pm0.7$ & $19.4\pm0.7$ \\
& lognorm fit $\mathrm{Mo}_{\Delta t_1}$& $15.3\pm0.9$& $15.4\pm0.9$ &
  $15.3\pm0.9$ & $15.3\pm0.9$ & $15.4\pm0.9$ \\ \hline
\multirow{2}{2em}{} & \textit{t}-test full dataset & $8.4$ & $6.6$ & $6.8$
  & $8.7$ & $7.6$ \\
& av.\ \textit{t}-score subset & $2.9$ & $1.7$ & $1.7$ & $2.3$ & $2.6$ 
\\ \hline \end{tabular}
\end{minipage}
\end{table}

\begin{table}[h!]
\renewcommand*{\arraystretch}{1.5}
\caption{\label{tab:selfResults}%
Distributional properties of the time lag until first citation when this is
a self-citation. The mean ($\overline{\Delta t_1}$), median
($\widetilde{\Delta t_1}$), and modal ($\mathrm{Mo}_{\Delta t_1}$) values
derived directly from the empirically observed distribution are shown
alongside corresponding values obtained when the high peak at $\Delta t_1 =
0$ is replaced by the average of probabilities observed for $\Delta t_1 =
\pm 2\, $months or completely removed.}
\begin{tabular}{|l|c|c|c|c|c} \hline
{} & $\overline{\Delta t_1}$ & $\widetilde{\Delta t_1}$ &
$\mathrm{Mo}_{\Delta t_1}$ \\ \hline
using all raw data from empirical distribution & $19.9$ & $17$ & $0$ \\
with $\Delta t_1 = 0$ peak adjusted by interpolation & $21.7\pm0.9$ &
$19.4\pm0.7$ & $15.0\pm0.9$ \\
with $\Delta t_1 = 0$ peak removed & $21.6\pm0.9$ & $19.4\pm0.7$ &
$15.2\pm0.9$ \\ \hline \end{tabular}
\end{table}

%
%\section*{Additional Files}
%  \subsection*{Additional file 1 --- Sample additional file title}
%    Additional file descriptions text (including details of how to
%    view the file, if it is in a non-standard format or the file extension).
%    This might refer to a multi-page table or a figure.
%
%  \subsection*{Additional file 2 --- Sample additional file title}
%    Additional file descriptions text.
%
\end{backmatter}
\end{document}